\documentclass[12pt]{article}
\pdfoutput=1
\usepackage{euscript,amsmath,amsthm,amsfonts,amssymb}
\usepackage{mathtools}
\usepackage{subfig}
\usepackage{float}
\usepackage[margin=1in]{geometry}
\usepackage{graphicx}
\usepackage{outlines}
\usepackage[dvipsnames]{xcolor}
\usepackage{jheppub}
\usepackage{cleveref}
\newcommand{\ket}[1]{|{#1}\rangle}

\newcommand{\bra}[1]{\langle{#1}|}

\DeclareMathOperator{\Tr}{Tr}

\title{\boldmath Clifford operators in $SU(N)_1$; $N$ not odd prime}
\author{Howard J. Schnitzer}
\affiliation{Martin Fisher School of Physics, Brandeis University, Waltham, Massachusetts 02453, USA}
\preprint{BRX-TH-6668}
\emailAdd{schnitzr@brandeis.edu}
\abstract{Farinholt gives a characterization of Clifford operators for qudits; $d$ both odd and even. In this comment it is shown that the necessary gates for the construction of Clifford operators; $N$ both odd and even, are obtained directly from operations that appear in $SU(N)_1$. A witness for $W_3$ states in $SU(2)_1$ is discussed. See e.g. \cite{Farinholt_2014,Schnitzer:2019icr,Eastin_2013,O_Gorman_2017}.}

\begin{document}
\maketitle
\flushbottom

\section{Introduction}
In applications there is a strong preference for qudits with $d$ prime, in the construction of the Pauli group and Clifford operators. This is exemplified by applications of $SU(N)_1$; $N$ prime and it's level-rank dual $U(1)_N$. We show, following Farinholt \cite{Farinholt_2014}, that the restriction to $N$ prime is not necessary for $SU(N)_1$ in the construction of the Pauli group and Clifford operators. The necessary operators are obtained from $SU(N)_1$.

\section{$SU(d)_1$ Pauli group}
Representations of $SU(d)_1$\footnote{In what follows we denote the group as $SU(d)_1$ rather than $SU(N)_{1}$ to describe qudits.} can be described by a single column Young tableau, with zero, one, ..., ($d$-1) boxes. The fusion tensor of the theory is
\begin{equation}\label{eq:1}
N_{ab}^c; \; a+b=c\mod d
\end{equation}
so that
\begin{equation}\label{eq:2}
N\ket{a}\ket{b}=\ket{a}\ket{a+b\mod d}.
\end{equation}
The modular transformation matrix $S_{ab}$ satisfies
\begin{equation}
\ket{a}=\sum_{b=0}^{d-1}S_{ab}\ket{b},\;a=0\text{ to }d-1.
\end{equation}
Let $\omega$ be a primitive $d$-th root of unity
\begin{equation}
\omega=\exp(\frac{2\pi i}{d})
\end{equation}
then it can be shown \cite{Schnitzer:2020lcr,Schnitzer:2019icr}
\begin{equation}\label{eq:5}
S^{*}=\frac{1}{\sqrt{d}}\sum_{a=0}^{d-1}\sum_{b=0}^{d-1}\omega^{ab}\ket{a}\bra{b}
\end{equation}
which is the $d$-dimensional generalization of the Hadamard gate. Equation \eqref{eq:5} can be rewritten as
\begin{equation}\label{eq:6}
S^{*}\ket{a} =\frac{1}{\sqrt{d}}\sum_{b=0}^{d-1}\omega^{ab}\ket{b}
\end{equation}
which is the $d$-dimensional discrete Fourier transform (QFT). With these ingredients, one can construct the qudit Pauli group.

\paragraph{$n=1$ qudits}\mbox{}\\

Let
\begin{equation}
Z_{ae}=\sum_{a,b=0}^{d-1}S_{cb}N^c_{b,1}(S^{\dagger}_{c,e})
\end{equation}
so that with \eqref{eq:1}-\eqref{eq:6},
\begin{equation}
Z_{ac}=\sum_{b=0}^{d-1}S_{ab}(S^{\dagger}_{b+1,a})\delta_{ac}
\end{equation}
or
\begin{equation}
Z=\sum_{a,b=0}^{d-1}S_{ab}(S^{\dagger}_{b+1,a})\ket{a}\bra{a},
\end{equation}
i.e.
\begin{equation}\label{eq:10}
Z=\sum^{d-1}_{a=0}\omega^a\ket{a}\bra{a},
\end{equation}
and
\begin{equation}\label{eq:11}
Z\ket{a}=\omega^a\ket{a},
\end{equation}
which is the Pauli $Z$. The modular transformation matrix is identical with the Pauli $X$, since
\begin{equation}
N_{a,1}^b\ket{a}=\ket{a+1,\mod d},
\end{equation}
which is identical to
\begin{equation}\label{eq:13}
X\ket{a}=\ket{a+1,\mod d},
\end{equation}
or
\begin{equation}\label{eq:14}
X=\ket{a+1}\bra{a}\mod d
\end{equation}
Therefore \eqref{eq:11} and \eqref{eq:14} are the basic ingredients for the single qudit Pauli group. From \eqref{eq:10} and \eqref{eq:14} 
\begin{equation}
(XZ)^r=\omega^{r(r-1)2}XZ
\end{equation}
when $d$ is odd $XZ$ has order $d$, and when $d$ is even $XZ$ has order $2d$. Define \cite{Farinholt_2014} $\hat{\omega}$ the primitive $D$-th root of unity where
\begin{equation}
\begin{split}
D=d;\;d\text{ odd}\\
D=2d;\;d\text{ even}
\end{split}
\end{equation}
The single qudit Pauli group is the collection of operators
\begin{equation}
\hat{\omega}^rX^aZ^b;\;r\in\mathbb{Z}_{D},\;a,b\in\mathbb{Z}_d.
\end{equation}
\begin{equation}
(X^aZ^b)(X^{a'}Z^{b'})=\omega^{ab'-ba'}(X^{a'}Z^{b'})(X^aZ^b),
\end{equation}
where the exponent of $\omega$ is identified with a symplectic product.

Thus all elements of the one-qudit Pauli group are obtained from basic operators of $SU(d)_1$
\paragraph{n-qudits}\mbox{}\\

Up to a global phase  \cite{Farinholt_2014}
\begin{equation}
X^{\underline{a}}Z^{\underline{b}}=X^{a_1}Z^{b_1}\otimes X^{a_2}Z^{b_2}\otimes ... \otimes X^{a_n}Z^{b_n}
\end{equation}
where
\begin{equation}
\underline{a}=(a_1,a_2,...,a_n)
\end{equation}
and
\begin{equation}
\underline{a}=(b_1,b_2,...,a_n)
\end{equation}
so that
\begin{equation}\label{eq:22}
(X^{\underline{a}}Z^{\underline{b}})(X^{\underline{a}'}Z^{\underline{b}'})=\omega^{(\sum^n_{i=1}a_ib'_i-a_i'b_i)}(X^{\underline{a}'}Z^{\underline{b}'})(X^{\underline{a}}Z^{\underline{b}}).
\end{equation}
Consider the operator $X^{\underline{a}}Z^{\underline{b}}$ along with all scalar multiples there of, where
\begin{equation}
\{\hat{\omega}^cX^{\underline{a}}Z^{b}|c\in \mathbb{Z}_D\}
\end{equation}
defines the n-qudit Pauli group. From \eqref{eq:22} this is isomorphic to the $2n$ commutative ring
\begin{equation}\label{eq:24}
M_{R}=\mathbb{Z}_D\times\mathbb{Z}_D\times...\times\mathbb{Z}_D.
\end{equation}
Multiplication in the Pauli group then corresponds to ring multiplication in \eqref{eq:24}.

Again all elements of the $n$ qudit Pauli group are obtained from direct products of basic operators of $SU(d)_1$. There ingredients allow one to construct $n$ qudit Clifford operators following Farinholt \cite{Farinholt_2014}.

\subsection*{$SU(d)_{1}$ Clifford operators}
\paragraph{Single-qudit Clifford operators  \cite{Farinholt_2014,1999}}\mbox{}\\

The necessary gates are 
\begin{outline}
\1[i)] The QFT gate \eqref{eq:6}
\1[ii)] The phase gate
\begin{equation}\label{eq:25}
\overline{P}\ket{j}=\omega^{\frac{j(j-1)}{2}}\ket{j},\quad j\text{ odd}
\end{equation}
\begin{equation}
\overline{P}\ket{j}=\omega^{\frac{j^2}{2}}\ket{j},\quad j\text{ even}
\end{equation}
which alternatively can be written as
\begin{equation}\label{eq:27}
\overline{P}\ket{j}=Z^{\frac{j(j-1)}{2}}\ket{j},\quad j\text{ odd}
\end{equation}
\begin{equation}
\overline{P}\ket{j}=\omega^{\frac{j}{2}}Z^{\frac{j(j-1)}{2}}\ket{j},\quad j\text{ even}.
\end{equation}
\end{outline}
\paragraph{Multi-qudit Clifford operators  \cite{Farinholt_2014,1999}}\mbox{}\\

The QFT and phase-gate are obtained from the natural product generalization of \eqref{eq:6} and \eqref{eq:25} - \eqref{eq:27}. One also needs the sum gate for a $n$-qudit system, with $i$ as the control and $j$ as the target qudit. From \eqref{eq:2} \cite{Farinholt_2014,1999}
\begin{equation}
\begin{split}
C_{\text{sum}}\ket{i}\ket{j}&=N\ket{i}\ket{j},\quad d\text{ odd}\\
&=\ket{i}\ket{i+j,\mod d}
\end{split}
\end{equation}
\begin{equation}
\begin{split}
C_{\text{sum}}\ket{i}\ket{j}&=\omega^{\frac{1}{2}(i+j)}N\ket{i}\ket{j},\quad d\text{ even}\\
&=\omega^{\frac{1}{2}(i+j)}\ket{i}\ket{i+j,\mod d}
\end{split}
\end{equation}

\paragraph{Toffeli gate \cite{1999,Ionicioiu_2009,Jones_2013,Eastin_2013,O_Gorman_2017,Biswal_2019,heyfron2019quantum}}\mbox{}\\

\begin{equation}
T^{(3)}\ket{i,j,k} =N^{(ij+k)}_{ij,k}=\ket{i,j;ij+k}\mod d, \quad d\text{ odd}
\end{equation}
from equation \eqref{eq:2}, while
\begin{equation}
T^{(3)}\ket{i,j,k} =\omega^{\frac{1}{2}(ij+k)}N^{(ij+k)}_{ij,k}, \quad d\text{ even}
\end{equation}

\paragraph{Multi-Toffeli gate}
\begin{equation}
\begin{split}
T^{(n)}\ket{a_1,a_2,...,a_{n-1},b}&=N^{(a_1,a_2,...,a_{n-1}+b)}_{a_1,a_2,...,a_{n-1},b}\\&=\ket{a_1,a_2,...,a_{n-1};a_1,a_2,...,a_{n-1}+b}\mod d,\quad d\text{ odd}
\end{split}
\end{equation}
\begin{equation}\label{eq:34}
T^{(n)}\ket{a_1,a_2,...,a_{n-1},b}=\omega^{\frac{1}{2}(a_1,a_2,...,a_{n-1}+b)}N^{(a_1,a_2,...,a_{n-1}+b)}_{a_1,a_2,...,a_{n-1},b},\quad d\text{ even}
\end{equation}
Equations \eqref{eq:25}-\eqref{eq:34} provide the resources for fault-tolerant computation for both $d$ odd and even.

\section{$W_3$ states are magical}

$W_3$ is magical by definition, since it is not a stabilizer state. The discussion of magic states for qubits is limited by the absence of the discrete Wigner function for qubits. However, there exist entanglement witnesses \cite{G_hne_2009} with non-local stabilizing operators which can detect three qubits states which are close to a $\ket{W_3}$ state,
\begin{equation}
\ket{W_3}=\frac{1}{\sqrt{3}}(\ket{100}+\ket{010}+\ket{001}),
\end{equation}
which is not a stabilizer state. A witness for this state is \cite{G_hne_2009}
\begin{equation}\label{eq:3.2}
\tilde{W}^{(W_3)}=\frac{2}{3}\mathbb{I}-\ket{W_3}\bra{W_3}
\end{equation}
Any witness for a $\ket{W_3}$ state has the property that
\begin{equation}
\Tr(\rho \mathcal{W})<0
\end{equation}
for a state which is close to $\ket{W_3}$. Therefore from \eqref{eq:3.2} one considers
\begin{equation}
\Tr(\rho \tilde{W}^{W_3})=\frac{2}{3}-\Tr(\rho\rho_{W_3})<0
\end{equation}
for states normalized to $\Tr\rho=1$. In particular
\begin{equation}
\Tr(\rho_{W_3}\tilde{W}^{W_3)})=\frac{2}{3}-\Tr(\rho^2_{W_3})<0
\end{equation}
or
\begin{equation}
\Tr(\rho^2_{W_3})>\frac{2}{3}.
\end{equation}

Following T\'{o}th and G\"{u}hne \cite{G_hne_2009}, one can create $\ket{W_3}$ from $\ket{000}$ using unitary operator. The generators of the stabilizer for $\ket{000}$ are
\begin{equation}\label{eq:3.7}
S_k^{(\ket{000})}=Z^{(k)}; \quad k=1,2,3.
\end{equation}
One can stabilize $\ket{W_3}$ by
\begin{equation}\label{eq:3.8}
S_k^{(W_3)}=US_k^{\ket{000}}U^{\dagger}
\end{equation}
The $U$ is not unique but one choice is \cite{G_hne_2009}
\begin{equation}\label{eq:3.9}
U=\frac{1}{\sqrt{3}}[X^{(1)}Z^{(2)}+X^{(2)}Z^{(3)}+Z^{(1)}X^{(3)}]
\end{equation}
In \eqref{eq:3.7} and \eqref{eq:3.9}, the $X^{(i)}$ and $Z^{(i)}$ are the Pauli operators for 3-qubits, obtained as direct products of the Pauli operators \eqref{eq:11} and \eqref{eq:13}, and are constructed as operations in $SU(2)_1$. The generators of stabilizing operators, based on \eqref{eq:3.8} are \cite{G_hne_2009}
\begin{equation}
\begin{split}
S_1^{(W_3)}=\frac{1}{3}[Z^{(1)}+2Y^{(1)}Y^{(2)}Z^{(3)}+2X^{(1)}Z^{(2)}X^{(3)}]\\
S_2^{(W_3)}=\frac{1}{3}[Z^{(2)}+2Z^{(1)}Y^{(2)}Y^{(3)}+2X^{(1)}X^{(2)}Z^{(3)}]\\
S_3^{(W_3)}=\frac{1}{3}[Z^{(3)}+2Y^{(1)}Z^{(2)}Y^{(3)}+2Z^{(1)}X^{(2)}X^{(3)}]
\end{split}
\end{equation}
which are non-local. T\'{o}th and G\"{u}hne \cite{G_hne_2009} present other witnesses for $\ket{W_{3}}$.

Magic states can be distilled by Toffeli gates, such as those presented above, as operations in $SU(2)_1$. Akers and Rath \cite{Akers:2019gcv} have argued that  holographic CFT states require a large amount of tripartite entanglement. Witnesses will be helpful in pursuing that issue.

\section{Comments}
For $d$ prime, only a linear number of gates are needed to implement a Clifford operation in $d$-dimensional Hilbert space, while in general $\mathcal{O}(D\log D)$ are needed to implement a Clifford operator for $d$ even \cite{Farinholt_2014}. A strong preference for $d$ prime emerges in terms of the number of resources required to construct gates, using Clifford operations and stabilizer states, and for magic state models \cite{White:2020zoz,Bravyi_2005,Howard_2012,Campbell_2012,Veitch_2014,Vallone_2008,Ritz_2019,liu2020manybody}.

The comments of this note apply to Chern-Simons $SU(d)_1$ as well as its level-rank dual $U(1)_d$ \cite{Mlawer:1990uv}, which then extends Theorem 1 of \cite{Salton:2016qpp} to $d$ even.

\acknowledgments
We are grateful to Isaac Cohen and Jonathan Harper for their aid in preparing the manuscript.

\bibliographystyle{JHEP}
\bibliography{Howard_Clifford}
\end{document}